\documentclass[seceq]{ptptex}

\title{
Prepotential approach to solvable rational potentials \\
and exceptional orthogonal polynomials%
}

\author{
Choon-Lin \textsc{Ho}%
}

\inst{
Department of Physics, Tamkang University, Tamsui 25137, Taiwan,
R.O.C.}

\abst{
 We show how all the quantal systems related to the exceptional
Laguerre and Jacobi polynomials can be constructed in a direct and
systematic way, without the need of shape invariance and
Darboux-Crum transformation.  Furthermore, the prepotential need
not be assumed a priori.  The prepotential, the deforming
function, the potential, the eigenfunctions and eigenvalues are
all derived within the same framework.  The exceptional
polynomials are expressible as a bilinear combination of a
deformation function and its derivative.}

\begin{document}


\maketitle




\section{Introduction}

In the last three years or so one has witnessed some interesting
developments in the area of exactly solvable models in quantum
mechanics: the number of exactly solvable shape-invariant models
has been greatly increased owing to the discovery of new types of
orthogonal polynomials, called the exceptional $X_\ell$
polynomials.\cite{GKM1}\tocite{Que3} Unlike the classical
orthogonal polynomials, these new polynomials have the remarkable
properties that they still form complete sets with respect to some
positive-definite measure, although they start with degree $\ell$
polynomials instead of a constant.

Two families of such polynomials, namely, the Laguerre- and
Jacobi-type $X_1$ polynomials, corresponding to $\ell=1$, were
first proposed by G\'omez-Ullate et al. in Ref.~\citen{GKM1},
within the Sturm-Lioville theory, as solutions of second-order
eigenvalue equations with rational coefficients. The results in
Ref.~\citen{GKM1} were reformulated in the framework of quantum
mechanics and shape-invariant potentials by Quesne et al.
\cite{Que1,Que2}. These quantal systems turn out to be rationally
extended systems of the traditional ones which are related to the
classical orthogonal polynomials. The most general $X_\ell$
exceptional polynomials, valid for all integral $\ell=1,2,\ldots$,
were discovered by Odake and Sasaki \cite{OS1} (the case of
$\ell=2$ was also discussed in Ref.~\citen{Que2}). Later, in
Ref.~\citen{HOS} equivalent but much simpler looking forms of the
Laguerre- and Jacobi-type $X_{\ell}$ polynomials were presented.
Such forms facilitate an in-depth study of some important
properties of the $X_\ell$ polynomials, such as the actions of the
forward and backward shift operators on the $X_{\ell}$
polynomials, Gram-Schmidt orthonormalization for the algebraic
construction of the $X_{\ell}$ polynomials, Rodrigues formulas,
and the generating functions of these new polynomials. Structure
of the zeros of the exception polynomials was studied in
Ref.~\citen{HS}.

More recently, these exceptional polynomials have been studied in
many ways.  For instance, possible applications of these new
polynomials were considered in Ref.~\citen{MR} for
position-dependent mass systems, and in Ref.~\citen{Ho1} for the
Dirac and Fokker-Planck equations.  The new polynomials were also
considered as solutions associated with some conditionally exactly
solvable potentials \cite{DR}.  These polynomials were recently
constructed by means of the Darboux-Crum transformation
\cite{Que2,GKM2,STZ}. Rational extensions of certain
shape-invariant potentials related to the exceptional orthogonal
polynomials were generated by means of Darboux-B\"acklund
transformation in Ref.~\citen{Gran1}. Generalizations of
exceptional orthogonal polynomials to discrete quantum mechanical
systems were done in Ref.~\citen{OS2}.  Structure of the $X_\ell$
Laguerre polynomials was considered within the quantum
Hamilton-Jacobi formalism in Ref.~\citen{RPKKG}. Generalizations
of these new orthogonal polynomials  to multi-indexed cases were
discussed in Ref.~\citen{GKM4,OS3}. Recently radial oscillator
systems related to the exceptional Laguerre polynomials have been
considered based on higher order supersymmetric quantum mechanics
\cite{Que3}.

So far most of the methods employed to generate the exceptional
polynomials have invoked in one way or another the idea of shape
invariance and/or the related Darboux-Crum transformation.
Furthermore, the so-called superpotentials (which we shall call
the prepotential), which determine the potentials, have to be
assumed a priori (often with good educated guesses).

The aim of this paper is to demonstrate that it is possible to
generate all the quantal systems related to the exceptional
Laguerre and Jacobi polynomials by a simple constructive approach
without the need of shape invariance and the Darboux-Crum
transformation. The prepotential (hence the potential),
eigenfunctions and eigenvalues are all derived within the same
framework. We call this the prepotential approach, which is an
extension of the approach we employed to construct all the
well-known one-dimensional exactly solvable quantum potentials in
Ref.~\citen{Ho2}.

The plan of this paper is as follows.  Sect.~2 presents the ideas
of prepotential approach to systems which are rational extensions
of the traditional systems related to the classical orthogonal
polynomials.  In Sect.~3 the prepotential approach is employed to
generate the L2 Laguerre system. Construction of the J1 and J2
Jacobi cases is then outlined in Sect.~4. Sect.~5 summarizes the
paper.  Appendix A collects some useful results on the classical
Laguerre and Jacobi polynomials.  The L1 Laguerre system is then
summarized in Appendix B.

\section{Prepotential approach}

\subsection{Main ideas}

We shall adopt the unit system in which $\hbar$ and the mass $m$
of the particle are such that $\hbar=2m=1$. Consider a wave
function $\phi(x)$ which is written in terms of a function $W(x)$
as
\begin{eqnarray}
\phi (x)\equiv e^{W(x)}.
\end{eqnarray}
Operating on $\phi_N$ by the operator $-d^2/dx^2$ results in a
Schr\"odinger equation $\mathcal{H}\phi=0$, where
\begin{eqnarray}
\cal{H} &=&-\frac{d^2}{dx^2} + \bar{V},\\
\bar{V}&\equiv&  \dot{W}^{2} + \ddot{W}.
\end{eqnarray}
The dot denotes derivative with respect to $x$. Since $W(x)$
determines the potential $\bar{V}$, it is therefore called the
prepotential \cite{prepot}.  For clarity of presentation, we shall
often leave out the independent variable of a function if no
confusion arises.

In this work we consider the following form of the prepotential
\begin{eqnarray}
W(x,\eta) = W_0(x)-\ln \xi(\eta) + \ln p(\eta). \label{W}
\end{eqnarray}
Here $W_0(x)$ is the zero-th order prepotential, and $\eta(x)$ is
a function of $x$ which we shall choose to be one of the
sinusoidal coordinates, i.e., coordinates such that
$\dot{\eta}(x)^2$ is at most quadratic in $\eta$, since most
exactly solvable one-dimensional quantal systems involve such
coordinates. The choice of $\eta(x)$ and the final form of the
potential dictate the domain of the variable $x$. $\xi(\eta)$ and
$ p(\eta)$ are functions of $\eta$ to be determined later. We
shall assume $\xi(\eta)$ to be a polynomial in $\eta$.  The
function $p(\eta)$ consists of the eigen-polynomial, but itself
need not be a polynomial (see Sect. 3).

With the prepotential (\ref{W}), the wave function is
\begin{eqnarray}
\phi(x) =\frac{e^{W_0(x)}}{\xi(\eta)}\,p(\eta), \label{phi}
\end{eqnarray}
and the potential
 $\bar{V}=\dot{W}^2 + \ddot{W}$ takes the form
\begin{eqnarray}
\bar{V} &=& \dot{W}_0^{2}+\ddot{W}_0
+\left[\dot{\eta}^2\left(2\frac{\xi^{\prime
2}}{\xi^2}-\frac{\xi^{\prime
\prime}}{\xi}\right)-\frac{\xi^{\prime}}{\xi}\left(2\dot{W}_0\dot{\eta}
+\ddot{\eta}\right)\right]\nonumber\\
&&~~~~+\frac{1}{p}\left[\dot{\eta}^2p^{\prime\prime}+\left(2\dot{W}_0\dot{\eta}
+\ddot{\eta}-2\dot{\eta}^2\frac{\xi^{\prime
}}{\xi}\right)p^\prime\right].
 \label{V}
\end{eqnarray}
Here the prime denotes derivative with respective to $\eta$.

For $\xi(\eta)=1$, the prepotential approach can generate exactly
and quasi-exactly solvable systems associated with the classical
orthogonal polynomials \cite{Ho2}.  The presence of $\xi$ in the
denominators of $\phi(x)$ and $V(x)$ thus gives a rational
extension, or deformation, of the traditional system.  We
therefore call $\xi(\eta)$ the deforming function.

To make $\bar{V}$ exactly solvable, we demand that: (1) $W_0$ is a
regular function of $x$, (2) the deforming function $\xi(\eta)$
has no zeros in the the ordinary (or physical) domain of
$\eta(x)$, and (3) the function $p(\eta)$ does not appear in $V$.

The requirement (3) can be easily met by setting the last term
involving $p(\eta)$ in Eq.~(\ref{V}) to a constant, say
``$-\cal{E}$", i.e.,
\begin{eqnarray}
\dot{\eta}^2 p^{\prime\prime}+\left(2\dot{W}_0\dot{\eta}
+\ddot{\eta}-2\dot{\eta}^2\frac{\xi^{\prime
}}{\xi}\right)p^\prime+ {\cal{E}} p = 0.
 \label{p}
\end{eqnarray}
If $W_0$, $\xi$ can be determined, and Eq.~(\ref{p}) can be
solved, then we would have constructed an exactly solvable quantal
system $H\psi=\cal{E}\psi$ defined by $H =-d^2/dx^2 + V(x)$, with
the wave function (\ref{phi}) and the potential
\begin{eqnarray}
V(x)&\equiv& \dot{W}_0^{2}+\ddot{W}_0
+\left[\dot{\eta}^2\left(2\frac{\xi^{\prime
2}}{\xi^2}-\frac{\xi^{\prime
\prime}}{\xi}\right)-\frac{\xi^{\prime}}{\xi}\left(2\dot{W}_0\dot{\eta}
+\ddot{\eta}\right)\right].
 \label{Exact-V0}
\end{eqnarray}

\subsection{Determining $W_0(x), \xi(\eta)$ and $p(\eta)$}

As mentioned before, we require that $\xi(\eta)$ has no zeros in
the ordinary domain of the variable $\eta$, but $\xi(\eta)$ may
have zeros in the other region in the complex $\eta$-plane.
Suppose $\xi(\eta)$ satisfies the equation
\begin{eqnarray}
c_2(\eta)\xi^{\prime\prime} + c_1(\eta)\xi^{\prime} +
\widetilde{\mathcal{E}}\xi=0, ~~~\widetilde{\mathcal{E}}={\rm
real\  constant}.\label{xi-eq}
\end{eqnarray}
Here $c_2(\eta)$ and $c_1(\eta)$ are functions of $\eta$ to be
determined.  We want Eq.~(\ref{xi-eq}) to be exactly solvable.
This is most easily achieved by matching (\ref{xi-eq}) with the
(confluent) hypergeometric equation, and this we shall adopt in
this paper. Thus $c_2(\eta)$ and $c_1(\eta)$ are at most quadratic
and linear in $\eta$, respectively.

If the factor $\exp(W_0(x))/\xi(\eta)$ in Eq.~(\ref{phi}) is
normalizable, then $p(\eta)={\rm~ constant}$ (in this case we
shall take $p(\eta)=1$ for simplicity), which solves Eq.~(\ref{p})
with $\mathcal{E}=0$, is admissible.  This gives the ground state
\begin{eqnarray}
\phi_0(x)=\frac{e^{W_0(x)}}{\xi(\eta)}. \label{phi0}
\end{eqnarray}
However, if $\exp(W_0(x))/\xi(\eta)$ is non-normalizable, then
$\phi_0(x)$ cannot be the ground state. In this case, the ground
state, like all the excited states, must involve non-trivial
$p(\eta)\neq 1$. Typically it is in such situation that the
exceptional orthogonal polynomials arise. To determine states
involving non-trivial $p(\eta)$, we proceed as follows.
 Since both $\xi$ and
$\xi^\prime$ appear in Eq.~(\ref{p}), we shall take the ansatz
that $p(\eta)$ be a linear combination of $\xi$ and $\xi^\prime$:
\begin{eqnarray}
p(\eta)=\xi^\prime(\eta) F(\eta) + \xi(\eta)G(\eta),
\end{eqnarray}
where $F(\eta)$ and $G(\eta)$ are some functions of $\eta$. Then
using Eq.~(\ref{xi-eq}) we have
\begin{eqnarray}
p^\prime(\eta)=\xi^\prime\left[-\frac{c_1}{c_2}F+F^\prime+G\right]
+\xi\left[-\frac{\widetilde{\mathcal{E}}}{c_2}F+G^\prime\right].
\label{p-prime}
\end{eqnarray}
We demand that Eq.~(\ref{p}) be regular at the zeros of $\xi$.
This is achieved if $p^\prime\propto \xi$, which requires that the
coefficient of $\xi^\prime$ in Eq.~(\ref{p-prime}) be zero, thus
giving a relation that connects $F$ and $G$,
\begin{eqnarray}
G=\frac{c_1}{c_2}F-F^\prime.
 \label{G}
\end{eqnarray}
Putting Eq.~(\ref{G}) into (\ref{p}), we get
\begin{eqnarray}
&&\xi^\prime\left[-\dot{\eta}^2
\left(-\frac{\widetilde{\mathcal{E}}}{c_2}F+G^\prime\right)+\mathcal{E}F\right]\nonumber\\
+ &&\xi\left[\dot{\eta}^2\frac{d}{d\eta}
\left(-\frac{\widetilde{\mathcal{E}}}{c_2}F+G^\prime\right) +
\left(2\dot{W}_0\dot{\eta}
+\ddot{\eta}\right)\left(-\frac{\widetilde{\mathcal{E}}}{c_2}F+G^\prime\right)+\mathcal{E}G\right]=0.
\label{p1}
\end{eqnarray}
Since $\xi$ and $\xi^\prime$ are independent for any polynomial
$\xi$, Eq.~(\ref{p1}) implies the coefficients of $\xi$ and
$\xi^\prime$ are zero. Setting the terms in the two
square-brackets to zero, and using Eq.~(\ref{G}) to eliminate $G$,
we arrive at the following equations
 satisfied by $F(\eta)$ and $c_1(\eta)$, respectively:
\begin{eqnarray}
-\dot{\eta}^2 F^{\prime\prime} + \frac{\dot{\eta}^2}{c_2} c_1
F^\prime +
\frac{\dot{\eta}^2}{c_2}\left[c_2\frac{d}{d\eta}\left(\frac{c_1}{c_2}\right)-\widetilde{\mathcal{E}}\right]
F=\mathcal{E}F,\label{F}
\end{eqnarray}
and
\begin{eqnarray}
c_1(\eta)&=&
\frac{c_2}{\dot{\eta}^2}\left[\frac{d}{d\eta}\left(\dot{\eta}^2\right)-
\left(2\dot{W}_0\dot{\eta} +\ddot{\eta}\right)\right]\nonumber\\
&=&\frac{c_2}{\dot{\eta}^2}\left[\frac12\frac{d}{d\eta}\left(\dot{\eta}^2\right)-
2Q(\eta)\right], \label{c1}
\end{eqnarray}
where $Q(\eta)\equiv \dot{W}_0\dot{\eta}$, and we have used the
identity $\ddot{\eta}=(d\dot{\eta}^2/d\eta)/2$ to arrive at the
last line.

Eq.~(\ref{G}) suggests that we set
\begin{equation}
F(\eta)=c_2(\eta) \cal{V}(\eta), \label{F-V}
\end{equation}
with some function $\mathcal{V}(\eta)$ in order to avoid any
possible singularity from $c_2$. Eqs.~(\ref{F}) and (\ref{G}) then
reduce to
\begin{eqnarray}
c_2\mathcal{V}^{\prime\prime} +
\left(2c_2^\prime-c_1\right)\mathcal{V}^\prime
+\left[c_2^{\prime\prime}-c_1^\prime+\widetilde{\mathcal{E}}+\frac{c_2}{\dot
{\eta}^2}\mathcal{E}\right]\mathcal{V}=0,
 \label{cal-V}
\end{eqnarray}
and
\begin{eqnarray}
G(\eta)=\left(c_1-c_2^\prime\right){\cal{V}}-c_2{\cal V}^\prime.
\label{G-1}
\end{eqnarray}
As mentioned before, in this paper we shall take $c_2(\eta)$ and
$c_1(\eta)$ to be at most quadratic and linear in $\eta$,
respectively. This means the coefficients of the first and second
terms in Eq.~(\ref{cal-V}) are also at most quadratic and linear
in $\eta$, respectively.  So Eq.~(\ref{cal-V}) can be matched with
the (confluent) hypergeometric equation, provided that the
coefficient of the last term in (\ref{cal-V}) is a constant. This
then requires
\begin{equation}
c_2(\eta)=\pm \dot{\eta}^2. \label{c2}
\end{equation}
(Note: in general one has $c_2(\eta)=\pm {\rm~constant} \times
\dot{\eta}^2$. But it is evident that the constant can be factored
out of Eq.~(\ref{cal-V}), together with a rescaling of
$\widetilde{\mathcal{E}}$. Thus without loss of generality we set
the constant to unity).  From Eq.~(\ref{c1}) this leads to
\begin{eqnarray}
c_1(\eta)=\pm\left[\frac12\frac{d}{d\eta}\left(\dot{\eta}^2\right)-
2Q(\eta)\right]. \label{c1-2}
\end{eqnarray}

Now we summarize the procedure or algorithm for constructing an
exactly solvable quantal system, whose potential as well as its
eigenfunctions and eigenvalues are all determined within the some
approach:

\begin{enumerate}
\item[$\bullet$] choose $\dot{\eta}^2$ from a sinusoidal coordinate; this
then fixes the form of  $c_2$;
\item[$\bullet$] by matching Eq.~(\ref{xi-eq}) with the (confluent) hypergeometric equation, one determines
$\tilde{\mathcal{E}},~Q(\eta)$, $c_1$ and $\xi(\eta)$. Integrating
$Q(x)=\dot{W}_0\dot{\eta}$ then gives the prepotential $W_0(x)$:
\begin{eqnarray}
W_0(x) &=& \int^x dx\, \frac{Q(\eta(x))}{\dot{\eta}(x)}\nonumber\\
&=& \int^{\eta(x)} d\eta\, \frac{Q(\eta)}{\dot{\eta}^2(\eta)};
\label{W0-fm-Q}
\end{eqnarray}
\item[$\bullet$] by matching Eq.~(\ref{cal-V}) with the (confluent) hypergeometric
equation, one determines $\mathcal{V}$, and thus
$F(\eta),~G(\eta),~p(\eta)$ and $\mathcal{E}$;
\item[$\bullet$] the exactly solvable system is defined by the wave function (\ref{phi}) and the potential
(\ref{Exact-V0}), which, by Eqs.~(\ref{xi-eq}) and (\ref{c2}), can
be recast in the form
\begin{eqnarray}
V(x)&\equiv& \dot{W}_0^{2}+\ddot{W}_0
+\frac{\xi^{\prime}}{\xi}\left[2\dot{\eta}^2\left(\frac{\xi^{\prime
}}{\xi}\right)-\left(2\dot{W}_0\dot{\eta} +\ddot{\eta}\right)\pm
c_1\right]\pm \tilde{\mathcal{E}}.
 \label{Exact-V}
\end{eqnarray}
\end{enumerate}

\subsection{Orthogonality of $p(\eta)$}

Using the relations
\begin{eqnarray}
\frac{d\xi}{d\eta}
=\frac{\dot{\xi}}{\dot{\eta}},~~\frac{dp}{dx}=\dot{\eta}p^\prime,~~
\frac{d^2p}{dx^2}=\dot{\eta}^2p^{\prime\prime}+
\ddot{\eta}p^\prime,
\end{eqnarray}
one can recast Eq.~(\ref{p}) into a differential equation in
variable $x$,
\begin{eqnarray}
\frac{d^2}{dx^2}p(\eta(x))
+2\left(\dot{W}_0-\frac{\dot{\xi}}{\xi}\right)\frac{d}{dx}p(\eta(x))
+\mathcal{E}p(\eta(x))=0.
\end{eqnarray}
This can further be put in the Sturm-Liouville form
\begin{eqnarray}
\frac{d}{dx}\left[\mathcal{W}^2\frac{d}{dx}p(\eta(x))\right]
+\mathcal{E}\mathcal{W}^2p(\eta(x))=0,
\end{eqnarray}
where
\begin{eqnarray}
\mathcal{W}(x) &\equiv &
\exp\left(\int^x dx\left(\dot{W}_0-\frac{\dot{\xi}}{\xi}\right)\right)\nonumber\\
&=& \frac{e^{W_0(x)}}{\xi(\eta(x))}. \label{cal-W}
\end{eqnarray}
According to the standard Sturm-Liouville theory, the functions
$p_{\mathcal{E}}$ (here we add a subscript to distinguish $p$
corresponding to a particular eigenvalue $\mathcal{E}$) are
orthogonal, i.e.,
\begin{equation}
\int dx~
p_{\mathcal{E}}(\eta(x))p_{\mathcal{E}'}(\eta(x))\mathcal{W}^2(x)\propto
\delta_{\mathcal{E},\mathcal{E'}}
\end{equation}
in the $x$-space, or
\begin{equation}
\int d\eta~
p_{\mathcal{E}}(\eta)p_{\mathcal{E}'}(\eta)\frac{\mathcal{W}^2(x(\eta))}{\dot{\eta}}\propto
\delta_{\mathcal{E},\mathcal{E'}} \label{ortho}
\end{equation}
in the $\eta$-space.

\section{L2 Laguerre case}

We now employ the above algorithm to generate the deformed radial
oscillator as given in Ref.~\citen{Que1,Que2,OS1}.

Let us choose $\eta(x)=x^2 \in [0,\infty)$. Then
$\dot{\eta}^2=4\eta$. For $c_2$ and $c_1$, we take the positive
signs in Eqs.~(\ref{c2}) and (\ref{c1-2}) (the opposite situation
is considered in Appendix B) . Thus $c_2(\eta)=4\eta$ and
$c_1=2(1-Q(\eta))$.

\subsection{$W_0,~\xi$ and $\tilde{\mathcal{E}}$}

Eq.~(\ref{xi-eq}) becomes
\begin{eqnarray}
\eta\xi^{\prime\prime} + \frac12 \left(1-Q(\eta)\right)\xi^\prime
+ \frac{\tilde{\mathcal{E}}}{4} \xi=0. \label{eq-xi-L2}
\end{eqnarray}
Comparing Eq.~(\ref{eq-xi-L2}) with the Laguerre equation
\begin{eqnarray}
\eta L_\ell^{\prime\prime(\alpha)} + \left(\alpha +1 -\eta
\right)L_\ell^{\prime(\alpha)} +\ell
L_\ell^{(\alpha)}=0,~~\ell=0,1,2,\ldots, \label{Lag}
\end{eqnarray}
where $L_\ell^{(\alpha)}(\eta)$ is the Laguerre polynomial, we
have
\begin{eqnarray}
\xi(\eta)\equiv
\xi_\ell(\eta;\alpha)=L_\ell^{(\alpha)}(\eta),~~\tilde{\mathcal{E}}=4\ell,
~~Q(\eta)=2\left(\eta-\alpha-\frac12\right). \label{xi-L2}
\end{eqnarray}
For $\xi_\ell(\eta;\alpha)$ not to have zeros in the ordinary
domain $[0,\infty)$, we must have $\alpha<-\ell$ (see Appendix A).
By Eq.~(\ref{W0-fm-Q}), the form of $Q(\eta)$ gives
\begin{eqnarray}
W_0(x)=\frac{x^2}{2}-\left(\alpha+\frac12\right)\ln x.
\label{W0-L2}
\end{eqnarray}
We shall ignore the constant of integration as it can be absorbed
into the normalization constant.

\subsection{$p(\eta),~\phi(\eta)$ and $\mathcal{E}$}

The above results implies that $\exp(W_0)\propto
\exp(x^2/2)x^{-(\alpha+\frac12)}$ ($\alpha<-\ell$).  The term
$\exp(x^2/2)$ will make $\phi(x)$ non-normalizable if $p(\eta)=1$,
or if $\mathcal{V}(\eta)$ is a polynomial in $\eta$. To remedy
this, we try $\mathcal{V}=\exp(-\eta)U(\eta)$ with some function
$U(\eta)$. Eq.~(\ref{cal-V}) becomes
\begin{eqnarray}
\eta U^{\prime\prime} + \left(-\alpha +1 -\eta \right)U^{\prime}
+\left(\frac{\mathcal{E}+\tilde{\mathcal{E}}}{4}+\alpha\right)U=0.
\end{eqnarray}
Comparing this equation with the Laguerre equation (\ref{Lag})
(replacing $\ell$ by another integer $n=0,1,2\ldots$. In the rest
of this paper, the index $n$ will always take on these values),
one has
\begin{eqnarray}
U(\eta)=L_n^{(-\alpha)}(\eta),~~\mathcal{E}\equiv
\mathcal{E}_n=4(n-\alpha)-\tilde{\mathcal{E}}=4(n-\alpha-\ell).
\label{L2-U-E}
\end{eqnarray}
From Eqs.~(\ref{F-V}) and (\ref{G-1}), one eventually obtains
\begin{eqnarray}
p(\eta)&\equiv& p_{\ell,n}(\eta)=\xi^\prime F+ \xi G\nonumber\\
&=&4e^{-\eta}P_{\ell,n}(\eta;\alpha),\\
P_{\ell,n}(\eta;\alpha)&\equiv&   \eta
L_n^{(-\alpha)}\xi_\ell^\prime + \left(\alpha L_n^{(-\alpha)}-\eta
L_n^{\prime(-\alpha)}\right)\xi_\ell\nonumber\\
&=& \eta L_n^{(-\alpha)}\xi_\ell^\prime + (\alpha-
n)L_n^{(-\alpha-1)}\xi_\ell. \label{P-L2}
\end{eqnarray}
Use has been made of Eqs.(\ref{L-1})-(\ref{L-3}) in obtaining the
last line in Eq.~(\ref{P-L2}). We note that
$P_{\ell,n}(\eta;\alpha)$ is a polynomial of degree $\ell+n$.  It
is just the L2 type exceptional Laguerre polynomial. We will show
in the next subsection that it is equivalent to the form presented
in Ref.~\citen{HOS} (to be called HOS form for simplicity).  By
Eq.~(\ref{ortho}), one finds that $P_{\ell,n}(\eta;\alpha)$'s are
orthogonal in the sense
\begin{equation}
\int^\infty_0\,d\eta~\frac{e^{-\eta}\eta^{-(\alpha+1)}}{\xi_\ell^2}\,
P_{\ell,n}(\eta;\alpha)P_{\ell,m}(\eta;\alpha)\propto \delta_{nm}.
\end{equation}

The exactly solvable potential is given by Eq.~(\ref{Exact-V})
with $W_0(x)$ and $\xi_\ell(\eta;\alpha)$ given by
Eqs.~(\ref{W0-L2}) and (\ref{xi-L2}), respectively.  The
eigenvalues $\mathcal{E}_n$ are given in Eq.~(\ref{L2-U-E}), i.e.
$\mathcal{E}_n=4(n-\alpha-\ell)$ .  Explicitly, the potential is
\begin{equation}
V(x)=x^2 +
\frac{\left(\alpha+\frac12\right)\left(\alpha+\frac{3}{2}\right)}{x^2}
+8\frac{\xi_\ell^{\prime}}{\xi_\ell}\left[\eta\left(\frac{\xi_\ell^{\prime}}{\xi_\ell}
-1\right)+\alpha+\frac12\right]+ 2(2\ell-\alpha).
\end{equation}
It is easily shown that $V(x)$ is equivalent to the potential for
L2 Laguerre case in Ref.~\citen{OS1,HOS,STZ} with
$\alpha=-g-\ell-\frac12~(g>0)$.  Particularly, it is exactly equal
to the form given in Eq.~(2.21) of Ref.~\citen{STZ}. The complete
eigenfunctions are
\begin{eqnarray}
\phi_{\ell,n}(x;\alpha) \propto
\frac{e^{-\frac{x^2}{2}}x^{-(\alpha+\frac12)}}{\xi_\ell}P_{\ell,n}(\eta;\alpha),~~\alpha
<-\ell.
\end{eqnarray}

For $\ell=0$, we have $\xi_0=1$ and $\xi_\ell^\prime=0$, and the
system reduces to the radial oscillator.  From Eq.~(\ref{P-L2})
one has $P_{\ell,n}\to L_n^{(-\alpha-1)}$ and $\alpha< -1/2$.

\subsection{Reducing $P_{\ell,n}(\eta;\alpha)$ to HOS form}

The polynomial $P_{\ell,n}(\eta;\alpha)$ is expressed as a
bilinear combination of $\xi_\ell(\eta;\alpha)$ and its derivative
$\xi_\ell^\prime (\eta;\alpha)$.  The HOS form instead expresses
the exceptional polynomial as a bilinear combination of
$\xi_\ell(\eta;\alpha)$ and its shifted form, i.e., $\xi_\ell
(\eta;\alpha-1)$.

To show the equivalence between $P_{\ell,n}(\eta;\alpha)$ and the
HOS form, we make use of the identities Eqs.~(\ref{L-1}) and
(\ref{L-3}) to express $\eta\xi_\ell^\prime (\eta;\alpha)$ in the
first term of $P_{\ell,n}(\eta;\alpha)$ as
\begin{eqnarray}
\eta\xi_\ell^\prime (\eta;\alpha)&=&-\eta
L_{\ell-1}^{(\alpha+1)}(\eta)\nonumber\\
&=&-\alpha L_{\ell-1}^{(\alpha)}(\eta)+\ell
L_\ell^{(\alpha-1)}(\eta).
\end{eqnarray}
Then we have
\begin{eqnarray}
P_{\ell,n}(\eta;\alpha) &=& \left(-\alpha
L_{\ell-1}^{(\alpha)}+\ell
L_\ell^{(\alpha-1)}\right)L_n^{(-\alpha)}+ \left(\alpha
L_n^{(-\alpha)}-\eta
L_n^{\prime(-\alpha)}\right)L_\ell^{(\alpha)}\nonumber\\
&=&\left(\alpha\left(
L_\ell^{(\alpha)}-L_{\ell-1}^{(\alpha)}\right)+\ell
L_\ell^{(\alpha-1)}\right)L_n^{(-\alpha)}-\eta
L_n^{\prime(-\alpha)}L_\ell^{(\alpha)}. \label{P-HOS}
\end{eqnarray}
Using Eq.~(\ref{L-2}) we have
$L_\ell^{(\alpha)}(\eta)-L_{\ell-1}^{(\alpha)}(\eta)=L_\ell^{(\alpha-1)}(\eta)$.
Finally, we arrive at
\begin{eqnarray}
P_{\ell,n}(\eta;\alpha)&=&\left(\alpha+\ell\right)
L_n^{(-\alpha)}(\eta)\xi_\ell(\eta;\alpha-1)-\eta
L_n^{\prime(-\alpha)}(\eta)\xi_\ell(\eta;\alpha),\label{L2-HOS}\\
&&~~~~~~~~~ \xi_\ell(\eta;\alpha-1) \equiv
L_\ell^{(\alpha-1)}(\eta).
\end{eqnarray}
Setting $\alpha=-g-\ell-\frac12$  and $\xi_\ell(\eta;g)\equiv
L_\ell^{(-g-\ell-\frac12)}(\eta)$ into (\ref{L2-HOS}), we have
\begin{eqnarray}
P_{\ell,n}(\eta;\alpha)&=&-\left[\left(g+\frac12\right)
L_n^{(g+\ell+\frac12)}(\eta)\,\xi_\ell(\eta;g+1)\right.\nonumber\\
&&\left.+\eta
L_n^{\prime(g+\ell+\frac12)}(\eta)\,\xi_\ell(\eta;g)\right].
\end{eqnarray}
This is, up to a multiplicative constant, the HOS form of the L2
Laguerre polynomial.

The example in this section demonstrates that the prepotential
approach described in Sect.~2 can indeed generate the exactly
solvable quantal system which has the L2 Laguerre polynomials as
the main part of its eigenfunctions.  The prepotential $W_0(x)$,
the potential $V(x)$, the deforming function
$\xi_\ell(\eta;\alpha)$, the eigenfunction
$\phi_{\ell,n}(x;\alpha)$ and eigenvalues $\mathcal{E}_n$ are all
determined from first principle.

In the next section and in the Appendix, we shall generate systems
associated with the exceptional Jacobi and L1 Laguerre
polynomials. Our description for these cases will be concise,
since the main steps are similar to those described in this
section.

\section{Exceptional Jacobi cases}

Let us choose $\eta(x)=\cos(2x)\in [-1,1]$. In this case it turns
out, as can be easily confirmed, that both the upper and the lower
signs in Eqs.~(\ref{c2}) and (\ref{c1-2}) for $c_2$ and $c_1$ give
the same equations that determine $\xi$ and $\mathcal{V}$, i.e.
Eqs.~(\ref{xi-eq}) and (\ref{cal-V}).  So for definiteness, we
shall take the upper signs, which give $c_2(\eta)= 4(1-\eta^2)$
and $c_1=- 2(2\eta+Q(\eta))$.

\subsection{$W_0,~\xi$ and $\tilde{\mathcal{E}}$}

Equation determining $\xi$ is
\begin{eqnarray}
(1-\eta^2)\xi^{\prime\prime}(\eta) +
\left(-\eta-\frac{Q(\eta)}{2}\right)\xi^\prime (\eta)+
\frac{\tilde{\mathcal{E}}}{4} \xi(\eta)=0. \label{eq-xi-J}
\end{eqnarray}
Comparing this with the differential equation satisfied by the
Jacobi polynomial $P_\ell^{(\alpha,\beta)}(\eta)$, namely,
\begin{equation}
  (1-\eta^2)P_\ell^{\prime\prime(\alpha,\beta)}(\eta)
  +\bigl(\beta-\alpha-(\alpha+\beta+2)\eta\bigr)P_\ell^{\prime(\alpha,\beta)}(\eta)
  +\ell(\ell+\alpha+\beta+1)P_\ell^{(\alpha,\beta)}(\eta)=0,
  \label{Jac}
\end{equation}
we have
\begin{eqnarray}
\xi(\eta)\equiv
\xi_\ell(\eta;\alpha,\beta)&=& P_\ell^{(\alpha,\beta)}(\eta),
~~\tilde{\mathcal{E}}=4\ell(\ell+\alpha+\beta+1),\nonumber\\
Q(\eta)&=&2\left[\alpha-\beta+\left(\alpha+\beta+1\right)\eta\right]
\label{xi-J}
\end{eqnarray}
for some parameters $\alpha$ and $\beta$. The form of $Q(\eta)$
gives, from Eq.~(\ref{W0-fm-Q}),
\begin{equation}
W_0(x)=-\left(\alpha + \frac12\right)\ln\sin x -\left(\beta +
\frac12\right)\ln\cos x. \label{W0-J}
\end{equation}
The equation of $\mathcal{V}$ is
\begin{equation}
(1-\eta^2)\mathcal{V}^{\prime\prime}
+\left[-\beta+\alpha-\left(-\beta-\alpha+2\right)\eta\right]\mathcal{V}^\prime
+\left(\frac{\mathcal{E}+\tilde{\mathcal{E}}}{4}+\alpha+\beta\right)\mathcal{V}=0.
\label{V-J}
\end{equation}

From Eq.~(\ref{W0-J}) we have
\begin{equation}
e^{W_0}\propto \left(1-\eta\right)^{-\frac12(\alpha+\frac12)}
\left(1+\eta\right)^{-\frac12(\beta+\frac12)}. \label{exp-W0-J}
\end{equation}
The exponents in Eq.~(\ref{exp-W0-J}) naturally divide the
parameters $\alpha$ and $\beta$ into four groups: (i)
$\alpha>-1/2,~\beta>-1/2$, (ii) $\alpha>-1/2,~\beta<-1/2$,  (iii)
$\alpha<-1/2,~\beta>-1/2$ and (iv) $\alpha<-1/2,~\beta<-1/2$.
Group (i) should be excluded, or $\xi_\ell(\eta;\alpha)$ will have
zeros in the ordinary domain $[-1,1]$ (see Appendix A).   So we
shall study the other three cases.  It turns out that these three
cases correspond, respectively, to quantal systems related to the
type J1, J2 exceptional Jacobi polynomials, and a rationally
extended Jacobi system obtained from the Darboux-P\"oschl-Teller
system by deleting the lowest $\ell$ excited states according to
the Crum-Adler method discussed in Ref.~\citen{GOS}.  We stress
here that the actual admissible parameters $\alpha$ and $\beta$ in
each case are dictated by the final form of $\mathcal{V}$, as will
be shown below.

\subsection{J1 Jacobi case}

Consider the case with parameters $\alpha>-1/2,~\beta<-1/2$. The
deforming function $\xi_\ell(\eta;\alpha,\beta)$ is given in
Eq.~(\ref{xi-J}),
\begin{equation}
\xi_\ell(\eta;\alpha,\beta)=P_\ell^{(\alpha,\beta)}(\eta).
\end{equation}
We demand that  $\xi_\ell(\eta;\alpha,\beta)$ has no zeros in the
ordinary domain $[-1,1]$. From Eq.~(\ref{no-zero-cond}), one can
easily check that this is the case if $\beta<-\ell$ for
$\alpha>-1/2$.  For this choice of the parameters the first term
$(1-\eta)^{-\frac12(\alpha+\frac12)}$ of Eq.~(\ref{exp-W0-J}) will
make the eigenfunction $\phi(x)$ non-normalizable, if
$\mathcal{V}$ is a polynomial.

This prompted us to try $\mathcal{V}=(1-\eta)^\gamma U(\eta)$
where $\gamma$ is a real parameter and $U(\eta)$ a function of
$\eta$. From Eq.~(\ref{V-J}) we find that $U(\eta)$ satisfies
\begin{eqnarray}
&&(1-\eta^2) U^{\prime\prime} +\left(-2\gamma-\beta+\alpha
-(2\gamma-\beta-\alpha+2)\eta \right)U^{\prime}\nonumber\\
&+& \left(\frac{\mathcal{E}+\tilde{\mathcal{E}}}{4}+\alpha +\beta+
\gamma(\gamma+\beta-\alpha-1)+
2\gamma(\gamma-\alpha)\frac{\eta}{1-\eta}\right)U=0. \label{U-J1}
\end{eqnarray}
If $\gamma=0,~\alpha$, the coefficient of $U$ in the last term of
the above equation can be reduced to a constant, so that
Eq.~(\ref{U-J1}) can be compared with the Jacobi differential
equation (\ref{Jac}). As $\gamma=0$ does not solve our original
problem with normalizability of the wave function, so we shall
take $\gamma=\alpha$.  This leads to
\begin{eqnarray}
(1-\eta^2) U^{\prime\prime} + \left(-\beta-\alpha
-(-\beta+\alpha+2)\eta \right)U^{\prime}
+\left(\frac{\mathcal{E}+\tilde{\mathcal{E}}}{4}+\beta(\alpha+1)\right)U=0.
\end{eqnarray}
Comparing this equation with Eq.~(\ref{Jac}), we conclude that
\begin{eqnarray}
&&~~~~~~~~~~~~~~~U(\eta;\alpha,\beta)=P_n^{(\alpha,-\beta)}(\eta),\nonumber\\
\mathcal{E}&&\equiv \mathcal{E}_n=4\left[n(n+\alpha-\beta +1)
-\ell(\ell+\alpha+\beta+1)-\beta(\alpha+1)\right]. \label{J1-U-E}
\end{eqnarray}
Putting all these results into $F(\eta)$ and $G(\eta)$ gives
\begin{eqnarray}
p(\eta)&\equiv&
p_{\ell,n}(\eta;\alpha,\beta)=4(1-\eta)^{\alpha+1}P_{\ell,n}(\eta;\alpha,\beta),\nonumber\\
P_{\ell,n}(\eta;\alpha,\beta)&\equiv& \left\{
(1+\eta)P_n^{(\alpha,-\beta)}(\eta)\xi_\ell^\prime +\left[\beta
P_n^{(\alpha,-\beta)}(\eta)-(1+\eta)P_n^{\prime(\alpha,-\beta)}(\eta)
\right]\xi_\ell\right\}\nonumber\\
&=& (1+\eta)P_n^{(\alpha,-\beta)}(\eta)\xi_\ell^\prime
-(n-\beta))P_n^{(\alpha+1,-\beta-1)}(\eta)\xi_\ell.\label{P-J1}
\end{eqnarray}
We have made use of Eq.~(\ref{J-6}) to arrive at the last line of
(\ref{P-J1}).  By Eq.~(\ref{ortho}), the orthogonality relations
of $P_{\ell,n}(\eta;\alpha)$'s are
\begin{equation}
\int^1_{-1}\,d\eta~\frac{(1-\eta)^{(\alpha+1)}(1+\eta)^{-(\beta+1)}}{\xi_\ell^2}\,
P_{\ell,n}(\eta;\alpha,\beta)P_{\ell,m}(\eta;\alpha,\beta)\propto
\delta_{nm}.
\end{equation}

The exactly solvable potential is given by Eq.~(\ref{Exact-V})
with $W_0(x)$ and $\xi_\ell(\eta;\alpha,\beta)$ given by
Eqs.~(\ref{W0-J}) and (\ref{xi-J}), respectively.  The eigenvalues
$\mathcal{E}_n$ are given in Eq.~(\ref{J1-U-E}).   The complete
eigenfunctions are
\begin{eqnarray}
\phi_{\ell,n}(x;\alpha,\beta) \propto
\frac{\left(1-\eta\right)^{\frac12(\alpha+\frac{3}{2})}
\left(1+\eta\right)^{-\frac12(\beta+\frac12)}}{\xi_\ell}
P_{\ell,n}(\eta;\alpha,\beta),\\
~~~~~~~~~\alpha>-1/2,~ \beta<-\ell.\nonumber
\end{eqnarray}
Using the identity (\ref{J-6}) one can show easily that
\begin{eqnarray}
P_{\ell,n}(\eta;\alpha,\beta)&=& (\ell+\beta)
P_n^{(\alpha,-\beta)}(\eta)\xi_\ell(\eta;\alpha+1,\beta-1)\nonumber\\
&-& (1+\eta)P_n^{\prime(\alpha,-\beta)}(\eta)
\xi_\ell(\eta;\alpha,\beta),
\end{eqnarray}
Up to a multiplicative constant, this is just the HOS form of the
J1 Jacobi polynomial presented in Ref.~\citen{HOS}, with the
substitution $\alpha=g+\ell-3/2$ and $\beta=-h-\ell-\frac12$. It
is easy to show that $V(x)$ and $\mathcal{E}_n$ are equivalent to
those for J1 Jacobi case given in Ref.~\citen{OS1,HOS,STZ} with
these values of $\alpha$ and $\beta$.

As $\ell\to 0$, the system reduces to the trigonometric
Darboux-P\"{o}schl-Teller potential, where $\alpha$ and $\beta$
can now take the values $\alpha>-3/2,~\beta<-1/2$.

\subsection{J2 Jacobi case}

One can proceed in a similar manner to construct the exactly
solvable systems with $\alpha<-1/2,~\beta>-1/2$.  This turns out
to lead to the system involving the  J2 Jacobi polynomials.

We shall not bore the readers with similar details here. Instead,
we point out  that it is easier to obtain the system by symmetry
consideration.  One notes that under the parity transformation
$\eta\to -\eta$, together with interchange $\alpha \leftrightarrow
\beta$, Eqs.~(\ref{eq-xi-J}) (with $Q(\eta)$ given by
(\ref{xi-J})) and (\ref{V-J}) are invariant in form.  This is in
complete accordance with the parity property of the Jacobi
polynomials, namely
\begin{equation}
  P_n^{(\alpha,\beta)}(-\eta)=(-1)^nP_n^{(\beta,\alpha)}(\eta).
  \label{Jparity}
\end{equation}
This implies that the J2 Jacobi system is simply the mirror image
of the J1 Jacobi system, and thus it can be obtained from the J1
case by taking the above transformations.

\subsection{Rationally extended Jacobi case}

Let $\alpha,~\beta <-1/2$.  In this case, the factors in
Eq.~(\ref{exp-W0-J}) cause no problem with normalizability of the
wave function even if $\mathcal{V}(\eta)$ is a polynomial. For
$\xi_\ell(\eta;\alpha,\beta)$ to be nodeless in the ordinary
domain $[-1,1]$, we must choose $\alpha$ and $\beta$ such that the
conditions in (\ref{no-zero-cond}) are satisfied. For example, if
$\ell=1$, one can have $\alpha<-1, ~-1<\beta<-1/2$, or $\beta<-1,
~-1<\alpha<-1/2$. For $\ell=2$, we have $\alpha,~\beta <-2$, or
$-2<\alpha,\,\beta<-1$.  For $\ell$ odd, we must have $\alpha\neq
\beta$, or $\xi(\eta)$ will have a zero at $\eta=0$ in view of
Eq.~(\ref{Jparity}).

Comparing Eqs.~(\ref{V-J}) and (\ref{Jac}), one obtains
\begin{eqnarray}
&&~~~~~~~\mathcal{V}(\eta)=P_n^{(-\alpha,-\beta)}(\eta),\nonumber\\
\mathcal{E}\equiv \mathcal{E}_n &=& 4\left[n(n-\alpha-\beta
+1)-\ell(\ell +\alpha +\beta +1) -\alpha-\beta\right].
\label{J-V-E-new}
\end{eqnarray}
From $F(\eta)$ and $G(\eta)$ we get
\begin{eqnarray}
&&p(\eta)\equiv P_{\ell,n}(\eta;\alpha,\beta)\equiv 4\left\{
(1-\eta^2)P_n^{(-\alpha,-\beta)}(\eta)\xi_\ell^\prime \right.\nonumber\\
&+&\left.\left[\left(\beta -
\alpha-\left(\beta+\alpha\right)\eta\right)
P_n^{(-\alpha,-\beta)}(\eta)-(1-\eta^2)P_n^{\prime(-\alpha,-\beta)}(\eta)
\right]\xi_\ell\right\}.
\end{eqnarray}
Again, by applying the identity (\ref{J-4}) and (\ref{J-6}), one
can reduce $P_{\ell,n}(\eta;\alpha,\beta)$ to
\begin{eqnarray}
&&P_{\ell,n}(\eta;\alpha,\beta)\nonumber\\
= &&4\left\{
(\ell+\beta)(1-\eta)P_n^{(-\alpha,-\beta)}(\eta)\xi_\ell(\eta;\alpha+1,\beta-1)\right.\nonumber\\
&+&\left.(n-\alpha)(1+\eta)P_n^{(-\alpha-1,-\beta+1)}(\eta)
\xi_\ell(\eta;\alpha,\beta)\right\}. \label{p-new}
\end{eqnarray}
One notes that $P_{\ell,n}(\eta;\alpha,\beta)$ is a polynomial of
degree $\ell+n+1$, and has $n+1$ nodes.  Thus the wave function
with $P_{\ell,0}(\eta;\alpha,\beta)$ has one node, and does not
correspond to the ground state. In fact, in this case the ground
state wave function is given by Eq.~(\ref{phi0}) with $p(\eta)=1$
and $\mathcal{E}=0$, since $\phi_0(x)$ is normalizable.  To ensure
that the energies of the excited states are positive, i.e.,
$\mathcal{E}_n>0$ for $n=0,1,2,\ldots$, one must have, besides the
constraints stated at the beginning of this subsection, the
condition $\alpha+\beta<-\ell$, which can be easily checked from
the form of $\mathcal{E}$ in Eq.~(\ref{J-V-E-new}).

The functions $P_{\ell,n}(\eta;\alpha,\beta)$ ($n=0,1,2,\ldots$),
together with $p(\eta)=1$, form a complete set and are orthogonal
with respect to the weight function
\begin{equation}
\frac{(1-\eta)^{-(\alpha+1)}(1+\eta)^{-(\beta+1)}}{\xi_\ell^2}.
\end{equation}
The complete eigenfunctions are given by
\begin{eqnarray}
\phi_0(x;\alpha,\beta) &\propto&
\frac{\left(1-\eta\right)^{-\frac12(\alpha+\frac12)}
\left(1+\eta\right)^{-\frac12(\beta+\frac12)}}{\xi_\ell},\nonumber\\
\phi_{\ell,n}(x;\alpha,\beta) &\propto&
\frac{\left(1-\eta\right)^{-\frac12(\alpha+\frac12)}
\left(1+\eta\right)^{-\frac12(\beta+\frac12)}}{\xi_\ell}
P_{\ell,n}(\eta;\alpha,\beta).
\end{eqnarray}
The exactly solvable potential is given by Eq.~(\ref{Exact-V})
with $W_0(x)$ and $\xi_\ell(\eta;\alpha,\beta)$  given by
Eqs.~(\ref{W0-J}) and (\ref{xi-J}), respectively. Since the
polynomials in the eigenfunctions start with degree zero, the
polynomials $ P_{\ell,n}(\eta;\alpha,\beta)$ cannot be considered
as exceptional.  In fact, this system corresponds to the system
discussed in Ref.~\citen{GOS}, which is obtained from the
Darboux-P\"oschl-Teller system by deleting the lowest $\ell$
excited states according to the Crum-Adler method.  It belongs to
the same class of rationally extended exactly solvable systems
discussed in Ref.~\citen{Gran2,Ho3}.  It is easy to show, using
the identities in Appendix A.2, that the polynomials in
Eq.~(\ref{p-new}) are proportional to those given by Eq.~(A.33) of
Ref.~\citen{GOS} for the Jacobi case.

\section{Summary}

We have demonstrated how all the quantal systems related to the
exceptional Laguerre and Jacobi  polynomials can be constructed in
a direct and systematic way.  In this approach one does not need
to rely on the requirement of shape invariance and the
Darboux-Crum transformation. Even the prepotential need not be
assumed a priori.  The prepotential, the deforming function, the
potential, the eigenfunctions and eigenvalues are all derived
within the same framework. It is worth to note that the main part
of the eigenfunctions, which are the exceptional orthogonal
polynomials, can be expressed as bilinear combination of the
deformation function $\xi(\eta)$ and its derivative
$\xi^\prime(\eta)$. However, they are equivalent of the forms
given in Ref.~\citen{HOS}.  We have also derived easily a
rationally extended Jacobi model obtained from the
Darboux-P\"oschl-Teller system by deleting the lowest $\ell$
excited states according to the Crum-Adler method discussed in
Rer.~\citen{GOS}.

We have not discussed the related hyperbolic
Darboux-P\"{o}schl-Teller systems (which are of J2 type).  They
can be generated in the same way by choosing the appropriate
sinusoidal coordinates. They can also be obtained from the
trigonometric case by suitable analytic continuation.

\section*{Acknowledgments}

This work is supported in part by the National Science Council
(NSC) of the Republic of China under Grant NSC
NSC-99-2112-M-032-002-MY3.



\appendix

\section{Useful identities}

In this Appendix, we collect some useful identities satisfied by
the Laguerre and Jacobi polynomials which are used in the main
text.

\subsection{Laguerre Polynomials}

Some useful relations among Laguerre polynomials are:
\begin{eqnarray}
\frac{d}{d\eta}L_\ell^{(\alpha)}(\eta)&=&-L_{\ell-1}^{(\alpha+1)}(\eta),
\label{L-1}\\
L_\ell^{(\alpha)}(\eta)+L_{\ell-1}^{(\alpha+1)}(\eta)&=&L_\ell^{(\alpha+1)}(\eta),
\label{L-2}\\
\eta L_{\ell-1}^{(\alpha+2)}(\eta) -\left(\alpha +1\right)
L_{\ell-1}^{(\alpha+1)}(\eta)&=&-\ell L_\ell^{(\alpha)}(\eta),
\label{L-3}
\end{eqnarray}

According to the Theorem 6.73 of Ref.~\citen{Szego}, for an
arbitrary real number $\alpha\neq-1,-2,\ldots,-\ell$, the number
of the positive zeros of $L_\ell^{(\alpha)}(\eta)$ is $\ell$ if
$\alpha>-1$; it is $\ell+[\alpha] +1$ if $-\ell<\alpha<-1$; it is
$0$ if $\alpha<-\ell$. Here $[a]$ denotes the integral part of
$a$. Furthermore, $\eta=0$ is a zero when and only when $\alpha=
-1,-2,\ldots,-\ell$.

\subsection{Jacobi polynomials}

Some useful relations among the Jacobi polynomial are:
\begin{eqnarray}
\frac{d}{d\eta}P_\ell^{(\alpha,\beta)}(\eta)&=&
\frac{\ell+\alpha+\beta+1}{2}P_{\ell-1}^{(\alpha+1,\beta+1)}(\eta),
\label{J-1} \\
2(\beta+1)
P_{\ell}^{(\alpha-1,\beta+1)}(\eta)&+&(\ell+\alpha+\beta+1)(\eta
+1)P_{\ell-1}^{(\alpha,\beta+2)}(\eta)\nonumber\\
&&~~~~~~~~~~~~=2(\ell+\beta+1)P_\ell^{(\alpha,\beta)}(\eta),
\label{J-2}\\
(\ell+\alpha)P_\ell^{(\alpha-1,\beta+1)}(\eta)&-&\alpha
P_{\ell}^{(\alpha,\beta)}(\eta)\nonumber\\
&&~=\frac{1}{2}(\ell+\alpha+\beta+1)(\eta-1)P_{\ell-1}^{(\alpha+1,\beta+1)}(\eta),
\label{J-3}\\
(n+\alpha)(1+\eta)P_{\ell-1}^{(\alpha,\beta+1)}(\eta)
  &-&\beta(1-\eta)P_{\ell-1}^{(\alpha+1,\beta)}(\eta)
  =2\ell P_\ell^{(\alpha,\beta-1)}(\eta).
  \label{J-4}
\end{eqnarray}

Using Eqs.~(\ref{J-1}) and (\ref{J-3}) to eliminate the
$P_{\ell-1}^{(\alpha+1,\beta+1)}(\eta)$ term gives
\begin{eqnarray}
\left(1-\eta\right)\frac{d}{d\eta}P_\ell^{(\alpha,\beta)}(\eta)
=\alpha
P_{\ell}^{(\alpha,\beta)}(\eta)-(\ell+\alpha)P_\ell^{(\alpha-1,\beta+1)}(\eta).
\label{J-5}
\end{eqnarray}
Combining Eqs.~(\ref{J-1}) and (\ref{J-2}) to eliminate the
$P_{\ell-1}^{(\alpha,\beta+2)}(\eta)$ term gives
\begin{eqnarray}
\left(1+\eta\right)\frac{d}{d\eta}P_\ell^{(\alpha-1,\beta+1)}(\eta)
=-(\beta+1)P_\ell^{(\alpha-1,\beta+1)}(\eta)+(\ell+\beta+1)
P_{\ell}^{(\alpha,\beta)}(\eta). \label{J-6}
\end{eqnarray}
Setting $\alpha\to \alpha+1,~\beta\to \beta-1$, we get
\begin{eqnarray}
\left(1+\eta\right)\frac{d}{d\eta}P_\ell^{(\alpha,\beta)}(\eta)
=(\ell+\beta) P_\ell^{(\alpha+1,\beta-1)}(\eta)-\beta
P_\ell^{(\alpha,\beta)}(\eta). \label{J-7}
\end{eqnarray}

According to the Theorem 6.72 of Ref.~\citen{Szego}, for arbitrary
real values of $\alpha$ and $\beta$, the number of zeros of
$P_\ell^{(\alpha,\beta)}(\eta)$ in  $(-1,1)$ is
\begin{eqnarray}
N_1(\alpha,\beta)
= \left\{
\begin{array}{ll}
  2\left[\frac{X+1}{2}\right],
  & ~~\text{if}~~ (-1)^\ell
  \left(\begin{array}{c}
  \ell+\alpha\\\ell
  \end{array}\right)
  \left(\begin{array}{c}
  \ell+\beta\\\ell
  \end{array}\right)>0;\\
  2\left[\frac{X}{2}\right]+1,
  & ~~\text{if}~~ (-1)^\ell
  \left(\begin{array}{c}
  \ell+\alpha\\\ell
  \end{array}\right)
  \left(\begin{array}{c}
  \ell+\beta\\\ell
  \end{array}\right)<0.
\end{array}\right.
\end{eqnarray}
Here
\begin{eqnarray}
X\equiv
E\left[\frac12\left(|2\ell+\alpha+\beta+1|-|\alpha|-|\beta|+1\right)\right],
\end{eqnarray}
where $E(u)$ is the Klein's symbol defined by
\begin{eqnarray}
E(u)=\left\{
\begin{array}{ll}
0,  &~~ u\leq 0; \nonumber\\
\left[u\right] &~~ u>0,~~u ~~\text{non-integral}; \\
u-1 &~~ u=1,2,3,\ldots\nonumber
\end{array}\right.
\end{eqnarray}

From this theorem, we conclude that the conditions for
$P_\ell^{(\alpha,\beta)}(\eta)$ to have no zeros in the ordinary
domain $(-1,1)$ are
\begin{eqnarray}
&&|2\ell+\alpha+\beta+1|-|\alpha|-|\beta|+1 \leq 0,\nonumber\\
\text{and} && (-1)^\ell
  \left(\begin{array}{c}
  \ell+\alpha\\\ell
  \end{array}\right)
  \left(\begin{array}{c}
  \ell+\beta\\\ell
  \end{array}\right)>0.\label{no-zero-cond}
\end{eqnarray}

It is noted that $\eta=+1 (-1)$ is a zero of
$P_\ell^{(\alpha,\beta)}(\eta)$ if and only if $\alpha
(\beta)=-1,-2,\ldots,-\ell$ with multiplicity
$|\alpha|~(|\beta|)$.

\section{L1 Laguerre case}

As with the L2 Laguerre case discussed in Sect.~3, let us take
$\eta(x)=x^2$. But now the negative signs in Eqs.~(\ref{c2}) and
(\ref{c1-2}) will be taken leading to $c_2(\eta)=-4\eta$ and
$c_1=-2(1-Q(\eta))$.

\subsection{$W_0,~\xi$ and $\tilde{\mathcal{E}}$}

Equation determining $\xi$ is
\begin{eqnarray}
-\eta\xi^{\prime\prime}(\eta) -\frac12
\left(1-Q(\eta)\right)\xi^\prime (\eta)+
\frac{\tilde{\mathcal{E}}}{4} \xi(\eta)=0. \label{eq-xi-L1}
\end{eqnarray}
We shall take $\mathcal{E}>0$, otherwise  the problem reduces to
the L2 Laguerre case discussed in Sect.~3.  The first term of
Eq.~(\ref{eq-xi-L1}) differs in sign from that of the the Laguerre
equation (\ref{Lag}).  Suppose we make a parity change $\eta\to
-\eta$ in Eq.~(\ref{eq-xi-L1}), then we will have
\begin{eqnarray}
\eta\xi^{\prime\prime}(-\eta) + \frac12
\left(1-Q(-\eta)\right)\xi^\prime (-\eta)+
\frac{\tilde{\mathcal{E}}}{4} \xi(-\eta)=0. \label{eq-xi-L1-2}
\end{eqnarray}
This equation has the form of Eq.~(\ref{Lag}), provided that
\begin{eqnarray}
\xi(-\eta)\equiv
\xi_\ell(-\eta;\alpha)=L_\ell^{(\alpha)}(\eta),~~\tilde{\mathcal{E}}=4\ell,~~
Q(-\eta)=2\left(\eta-\alpha-\frac12\right)
\end{eqnarray}
for some parameter $\alpha$.  This means
\begin{eqnarray}
\xi_\ell(\eta;\alpha)=L_\ell^{(\alpha)}(-\eta), \label{xi-L1}
\end{eqnarray}
and
\begin{equation}
Q(\eta)=-2\left(\eta+\alpha+\frac12\right).
\end{equation}
For $\xi_\ell(\eta;\alpha)$ not to have zeros in the ordinary
domain $[0,\infty)$, we must have $\alpha>-1$  at the least
(precise bound will be determined later). The form of $Q(\eta)$
then leads to
\begin{eqnarray}
W_0(x)=-\frac{x^2}{2}-\left(\alpha+\frac12\right)\ln x.
\label{W0-L1}
\end{eqnarray}
As before we ignore the constant of integration.

\subsection{$p(\eta),~\phi(\eta)$ and $\mathcal{E}$}

Consider $\exp(W_0)\propto \exp(-x^2/2)x^{-(\alpha+\frac12)}$
($\alpha>-1$ at the least).  Contrary to the L1 case, this time it
is the term $x^{-(\alpha+\frac12)}$ that could cause $\phi(x)$
non-normalizable (when $\alpha> -1/2$) if $\mathcal{V}(\eta)$ is a
polynomial in $\eta$. So we try $\mathcal{V}=\eta^\beta U(\eta)$
where $\beta$ is a real parameter and $U(\eta)$ a function of
$\eta$.  From Eq.~(\ref{cal-V}) we get
\begin{eqnarray}
\eta U^{\prime\prime} + \left(2\beta-\alpha +1 -\eta
\right)U^{\prime} +\left(\frac{\beta(\beta-\alpha)}{\eta} +
\frac{\mathcal{E} -\tilde{\mathcal{E}}}{4}-\beta-1\right)U=0.
\label{U-L1}
\end{eqnarray}
If $\beta=0,~\alpha$, the $\eta$-dependent term in the last term
of the above equation can be eliminated, and Eq.~(\ref{U-L1}) can
be reduced to the Laguerre equation (\ref{Lag}). As $\beta=0$ does
not solve our original problem with normalizability of the wave
function, we shall take $\beta=\alpha$.  This leads to
\begin{eqnarray}
U(\eta)=L_n^{(\alpha)}(\eta),~~\mathcal{E}\equiv
\mathcal{E}_n=4(n+\alpha+\ell+1). \label{L1-U-E}
\end{eqnarray}
Putting all these result into $F(\eta)$ and $G(\eta)$ gives
\begin{eqnarray}
p(\eta)&\equiv&
p_{\ell,n}(\eta)=-4\eta^{\alpha+1}P_{\ell,n}(\eta;\alpha)\nonumber\\
P_{\ell,n}(\eta;\alpha)&\equiv & L_n^{(\alpha)}\xi_\ell^\prime
+\left(L_n^{(\alpha)}-
L_n^{\prime(\alpha)}\right)\xi_\ell\nonumber\\
&=& L_n^{(\alpha)}\xi_\ell^\prime + L_n^{(\alpha+1)}\xi_\ell,
\label{P-L1}
\end{eqnarray}
where use has been made of Eqs.~(\ref{L-1}) and (\ref{L-2}) to get
the last line. $P_{\ell,n}(\eta;\alpha)$ is a polynomial of degree
$\ell+n$.  It will be shown below that it is just the L1 type
exceptional Laguerre polynomial.  It is also easy to check that
$P_{\ell,n}(\eta;\alpha)$'s are orthogonal with respect to the
weight function
\begin{equation}
\frac{e^{-\eta}\eta^{(\alpha+1)}}{\xi_\ell^2}.
\end{equation}

The exactly solvable potential is given by Eq.~(\ref{Exact-V})
with $W_0(x)$ and $\xi_\ell(\eta;\alpha)$ given by
Eqs.~(\ref{W0-L1}) and (\ref{xi-L1}), respectively.  The
eigenvalues are $\mathcal{E}_n=4(n+\alpha+\ell+1)$ . It is easy to
show that $V(x)$ is equivalent to the potential for L1 Laguerre
case in Ref.~\citen{OS1,HOS,STZ} with $\alpha=g+\ell-3/2~(g>0)$.
Particularly, it is exactly equal to the form of potential in
Eq.~(2.20) of Ref.~\citen{STZ}. The complete eigenfunctions are
\begin{eqnarray}
\phi_{\ell,n}(x;\alpha) \propto
\frac{e^{-\frac{x^2}{2}}x^{(\alpha+\frac{3}{2})}}{\xi_\ell}
P_{\ell,n}(\eta;\alpha),~~\alpha>-\frac{3}{2}.
\end{eqnarray}

As in the L2 case, this system reduces to the radial oscillator
system in the limit $\ell\to 0$.

\subsection{Reducing $P_{\ell,n}(\eta;\alpha)$ to HOS form}

Using Eqs.~(\ref{L-1}) and (\ref{L-2}), we have
\begin{eqnarray}
\xi_\ell^\prime (\eta;\alpha)=L_{\ell}^{(\alpha+1)}(-\eta)-
L_\ell^{(\alpha)}(-\eta).
\end{eqnarray}
Then it is easy to check that
\begin{eqnarray}
P_{\ell,n}(\eta;\alpha)&=&
L_n^{(\alpha)}(\eta)\xi_\ell(\eta;\alpha+1)-
L_n^{\prime(\alpha)}(\eta)\xi_\ell(\eta;\alpha)\label{L1-HOS}\\
&&~~~~~~~~~ \xi_\ell(\eta;\alpha+1) \equiv
L_\ell^{(\alpha+1)}(-\eta).
\end{eqnarray}
This is, up to a multiplicative constant, the HOS form of the L1
Laguerre polynomial, with the substitution $\alpha=g+\ell-3/2$.



\begin{thebibliography}{99}

\bibitem{GKM1}
D. G\'omez-Ullate, N. Kamran and R. Milson, J. Math. Anal.
Appl. {\bf 359} (2009) 352.\\
 D. G\'omez-Ullate, N. Kamran and R. Milson, J. Approx. Theory {\bf 162} (2010) 987.

\bibitem{Que1}
C. Quesne,  J. Phys. {\bf A41} (2008) 392001.\\
B.\,Bagchi, C.\,Quesne and R.\,Roychoudhury, Pramana J. Phys. {\bf
73} (2009) 337.

\bibitem{Que2}
C. Quesne, SIGMA {\bf 5} (2009) 084.

\bibitem{OS1}
S. Odake and R. Sasaki, Phys. Lett. {\bf B679} (2009) 414.\\
S. Odake and R. Sasaki, Phys. Lett. {\bf B684} (2009) 173. \\
S. Odake and R. Sasaki, J. Math. Phys. {\bf 51} (2010) 053513.

\bibitem{HOS}
C-L. Ho, S. Odake and R. Sasaki, ``Properties of the exceptional
($X_\ell$) Laguerre and Jacobi polynomials," YITP-09-70.
arXiv:0912.5447 [math-ph].

\bibitem{HS}
C-L. Ho and R. Sasaki, ``Zeros of the exceptional Laguerre and
Jacobi polynomials," Tamkang and YITP preprint, YITP-11-24, 2011.
arXiv: 1102.5669 [math-ph].

\bibitem{MR}
B. Midya and B. Roy, Phys. Lett. A {\bf 373} (2009) 4117.

\bibitem{Ho1}
C.-L. Ho, Ann. Phys. {\bf 326} (2011) 797.

\bibitem{DR}
D. Dutta and P. Roy, J. Math. Phys. {\bf 51} (2010) 042101.

\bibitem{GKM2}
D. G\'omez-Ullate, N. Kamran and R. Milson, J. Phys. {\bf A43}
(2010) 434016.

\bibitem{STZ}
R. Sasaki, S. Tsujimoto and A. Zhedanov, J. Phys. {\bf A43} (2010)
315204.

\bibitem{Gran1} Y. Grandati, ``Solvable rational extensions of the isotonic
oscillator."  arXiv:1101.0055 [math-ph].

\bibitem{OS2}
S.\,Odake and R.\,Sasaki,
Phys. Lett. {\bf B682} (2009) 130.\\
S.\,Odake and R.\,Sasaki, Prog. Theor.
Phys. {\bf 125} (2011) 851.\\
S.\,Odake and R.\,Sasaki, ``Discrete quatum mechanics,"
YITP-11-35. arXiv: 1104.0473 [math-ph].

\bibitem{RPKKG}
S.S. Ranjani, P.K. Panigrahi, A. Khare, A.K. Kapoor and A.
Gangopadhyaya, ``Exeptional orthogonal polynomials, QHJ formalism
and SWKB quantization condition". arXiv: 1009.1944 [math-ph].

\bibitem{GKM3} D.~G{\'o}mez-Ullate, N.~Kamran, and R.~Milson,
``On orthogonal polynomials spanning a non-standard
flag". arXiv:1101.5584 [math-ph].

\bibitem{GKM4}
D. G\'omez-Ullate, N. Kamran and R. Milson, ``Two-step Darboux
transformations and exceptional Laguerre polynomials". arXiv:
1103.5724 [math-ph].

\bibitem{OS3}
S.\,Odake and R.\,Sasaki, ``Exactly solvable quantum mechanics and
infinite families of multi-indexed orthogoanl polynomials".
arXiv:1105.0508 [math-ph].

\bibitem{Que3}
C. Quesne, ``Higher-order SUSY, exactly solvable potentials, and
exceptional orthogonal polynomials". arXiv:1106.1990 [math-ph].

\bibitem{Ho2} C.-L. Ho, Ann. Phys. {\bf 323} (2008) 2241.\\
C.-L. Ho, ``Prepotential approach to exact and quasi-exact
solvabilities of Hermitian and non-Hermitian Hamiltonians," (Talk
presented at ``Conference in Honor of CN Yang's 85th Birthday", 31
Oct - 3 Nov, 2007, Singapore).
arXiv:0801.0944 [hep-th].\\
C.-L. Ho, Ann. Phys. {\bf 324} (2009) 1095.\\
C.-L. Ho, J. Math. Phys. {\bf 50} (2009) 042105.\\
C.-L. Ho, Ann. Phys. {\bf 326} (2011) 1394.

\bibitem{prepot}
A.J. Bordner, N.S. Manton and R. Sasaki, Prog. Theor. Phys. {\bf
103} (2000) 463.\\
S.P. Khastgir, A.J. Pocklington and R. Sasaki, J. Phys. {\bf A33}
(2000) 9033.\\
E. Corrigan and R. Sasaki, J.Phys. {\bf A35} (2002) 7017.

\bibitem{GOS}
L. Garc\'ia-Guti\'errez, S, Odake and R. Sasaki, Prog. Theor.
Phys. {\bf 124} (2010) 1.

\bibitem{Gran2}
Y. Grandati, ``Solvable rational extensions of the Morse and
Kepler-Coulomb potentials". arXiv: 1103.5023 [math-ph].

\bibitem{Ho3}
C.-L. Ho, ``Prepotential approach to solvable rational extensions
of Harmonic Oscillator and Morse potentials". arXiv:1105.3670
[math-ph].

\bibitem{Szego}
G.\,Szeg\"o, Orthogonal Polynomials, Amer. Math. Soc. Colloquium
Publications Vol. 23 (Amer. Math. Soc., New York, 1939).

\end{thebibliography}
\end{document}